\documentstyle[12pt]{article}

\begin{document}

%%-Title and abstract page-%%

%%-move to normal A4-%%
\hoffset = -1truecm
\voffset = -2truecm

\title{\bf
Randall-Sundrum Zero Mode as a Penrose Limit 
}

\author{
{\bf
R. G{\" u}ven }\\
\normalsize Department of Mathematics, Bo\u{g}azi\c{c}i University, Bebek,
\.{I}stanbul 80815
{\bf Turkey}\\
\date{ }
}

\maketitle

\begin{abstract} A generalization of the limiting procedure of Penrose,
which allows non-zero cosmological constants and takes into account
metrics that contain homogeneous functions of degree zero, is presented.
It is shown that any spacetime which admits a spacelike conformal Killing
vector  has a limit which is conformal to plane waves. If the spacetime is an Einstein space, its limit exists only if the cosmological
constant is negative or zero. When the conformal Killing vector is
hypersurface orthogonal, the limits of Einstein  spacetimes are  certain AdS plane waves.
In this case the nonlinear version of the Randall-Sundrum zero mode is
obtained as the limit of the brane world scenarios.
 \end{abstract}

\newpage %%-main body of paper-%%
\newcommand{\be}{\begin{equation}}
\newcommand{\ee}{\end{equation}} 

The plane wave spacetimes are known to posssess various remarkable
properties in the context of both classical and quantum gravity. One of these is that any spacetime has a plane wave as its Penrose 
limit
\cite{kn:pen}. In a well defined sense plane waves play, around
null geodesics, a role that is similiar to the role of the tangent spaces about
the points of a manifold. This universal propery of the plane waves can be
seen by first choosing, on an arbitrary Lorentzian manifold, a suitable
coordinate gauge in a neighborhood of a null geodesic congruence. If the
neighborhood chosen in this manner is then blown up uniformly by rescaling
the appropriate coordinates, a plane wave spacetime results as the
limiting manifold.  Since the Einstein-Hilbert action transforms
homogeneously under the constant rescalings of the metric, the field
equations are preserved during the process. The limits of the vacuum
solutions therefore turn out to be the vacuum plane waves. The same
argument applies to all spacetime dimensions and extends to
Einstein-Maxwell fields. Moreover, in the context of the D=10 string
theories, the limiting procedure commutes with T-duality and the scaling
rules that are employed stem from the scaling propery of the D=11
supergravity action \cite{kn:guv}. Recent discussions of the role of the Penrose limit in the maximally supersymmetric
solutions, the worldvolume dynamics and the AdS/CFT correspondence can be found in \cite{kn:bl} \cite{kn:mal} \cite{kn:bbl}
 \cite{kn:itz} \cite{kn:gom} \cite{kn:russo} \cite{kn:son} \cite{kn:mat}.

       In its present form the limiting procedure, however, cannot accomodate a non-zero cosmological constant and does not take into
account the possibility of dealing with metrics that are homogeneous functions of degree zero in the coordinates. The purpose of this
paper is to furnish a generalization of the Penrose limit which covers such a situation. We shall consider spacetimes which admit
conformal Killing vectors and possess non-zero cosmological constants. It will be seen that a generalization of the Penrose limit exists
if the conformal Killing vector is spacelike and the cosmological constant is negative. Such spacetimes will have limits that are
conformal to the plane waves. If the conformal Killing vector is hypersurface orthogonal and the spacetime is an
Einstein space with a
negative cosmological constant, its limit is an anti-de Sitter (AdS) plane wave \cite{kn:sl}, \cite{kn:osv}, \cite{kn:cg}. From this
special case it
follows that the relevant
brane
world scenarios have the nonlinear version of the Randall-Sundrum zero
mode\cite{kn:rs},\cite{kn:cg} as a Penrose limit.

Consider a D-dimensional ($D\geq 4$)  spacetime $ M_D$ which admits a conformal Killing vector $K^{M}$~:
 \be {\mathcal{L}}_K g_{MN}=2\Phi g_{MN}, \ee
 where $  g_{MN}$ is the metric, $\Phi$ is a scalar field on $ M_D $ and $ \mathcal{L} $ is the Lie derivative .
Then one can always find another metric $\bar {g}_{MN}$ on a manifold $\bar{ M}_D$ which is conformal to $g_{MN}$:
 \be g_{MN} = W^{-2} \bar {g}_{MN},
 \ee
 for which $K^M$ is a Killing vector: ${\mathcal{L}_K} \bar {g}_{MN} = 0$. Using on $\bar{M_D}$ and $M_D$  the same coordinate patch $
X^M
=(x^{\mu}, Z)$  where $ \mu =0,
\ldots, D-2 $ and $ K^{M} = {\delta ^M}_Z$,  the standard Kaluza-Klein
decomposition of
$\bar{g}_{MN}$
gives
\be
ds^2 = W^{-2}(x^{\alpha}, Z)[\bar{g}_{\mu \nu}(x^{\alpha}) dx^{\mu}dx^{\nu} -{\lambda}^{2}(dZ + A_{\mu}(x^{\alpha})dx^{\mu})^2],
\ee
provided that $K^M$ is spacelike: $\bar {g}_{MN} K^M K^N = -{\lambda}^2 (x^{\alpha})$. Since the null geodesics are preserved under 
conformal transformations, we can choose in a conjugate point-free portion of a null geodesic congruence,  either on $\bar{M}_D$ or
$M_D$, the
Penrose coordinates\cite{kn:rose}: $ x^{\mu} = \{ Y^+, Y^-, Y^j\}$  where $ j= 1,\ldots,D-3$. In this coordinate system each null
geodesic
will be given by $ Y^{+} = const., Y^{j}=const.,
Z=const.$  with $ Y^+ $ labelling the different geodesics and $ Y^-$ is an affine parameter along the geodesics.
In this chart the line element $d{s_n}^2 =\bar{g}_{\mu \nu}dx^{\mu}dx^{\nu}$ with $n=D-1$ becomes 
\be
d{s_n}^2= 2dY^+ [ dY^- +  \alpha dY^+ + \beta_k dY^k ] - C_{jk} dY^j dY^k,
\ee
where the metric functions $\alpha, \beta_k$, $ C_{jk} $ depend in general on
all the
coordinates $x^{\mu}$ and $ C_{jk} $  is a $ (n-2) \times (n-2) $ positive definite symmetric matrix. Moreover, if we
let $A_{\mu}dx^{\mu} = A_{+}dY^{+} + A_{-}dY^{-} + A_{j}dY^{j}$, the Kaluza-Klein
gauge
field $ A_{\mu}$  obeys the gauge condition:
\be
A_{-} = 0.
\ee
Suppose now the conformal factor $W$ is decomposed as
\be
W(X^K) = B_M(X^K) X^{M}/{\it {l}} + w(X^K),
\ee
where $\it {l}$ is a real parameter and $ B_M(X^K)$ and $ w(X^K)$ are arbitrary. In the above coordinate gauge the function $ w$  will
be
chosen so that $B_{M}$ obeys
\be
B_{-} = 0,
\ee
in accordance with the gauge condition (5).

  After choosing the gauges in this manner let us rescale the coordinates by
introducing $ \{ U, V, X^j, \hat{Z} \} $ which satisfies
\be
 Y^- = U, \hspace{.3in} Y^+ =
{\Omega^2} V, \hspace{.3in} Y^j = {\Omega}X^j, \hspace{.3in} Z={\Omega}\hat{Z},
\ee
together with
\be
\it{l} = {\Omega}{\hat {\it {l}}},
\ee
where $\Omega > 0$ is a real number.
When these rescalings are taken into account in (3), one gets a two-parameter family of metrics
  $ {g}_{MN}(\Omega, \it{l})$ and it
 is useful to view these as  metrics on a two-parameter
family of spacetimes $ M_{D}(\Omega, \it{l}) $. This allows one to generalize the discussion given in \cite{kn:ger} and 
interpret the limit in a coordinate independent manner by
regarding
$\Omega$ and $\it{l}$ as  scalar fields on an associated (D  + 2) - dimensional manifold possessing a
degenerate metric and  boundary.  One boundary of the (D + 2) - dimensional
manifold is located at $\Omega =0$ and this will be the limit of interest. Before
approaching this boundary let us introduce on $ M_{D}(\Omega, \it{l})$ new
metrics
that are distinguished by hats and are related to the old ones by
 \be
{\hat{g}}_{MN}(\Omega, \hat{\it{l}}) = {\Omega ^{-2}}g_{MN}(\Omega, \it{l}).
 \ee
 Allowing now $\Omega \rightarrow 0$,
the rescaled line element (3)  becomes in the limit:
  \be d{\hat{s}}^2 = {W_{0}}^{-2} d{\bar{s}}^2,
 \ee
where
\be
 d{\bar{s}}^2 = 2dUdV - C_{jk}(U) dX^j dX^k - {\lambda}^{2}(U) (d\hat{Z} + A_{j}(U) dX^j)^2, 
\ee
is the standart plane wave metric, expressed in Rosen coordinates,  and assuming that $B_{M}(X^L)$ and $w(X^L)$ contain no piece which
is 
a homogeneous function of degree zero in $Y^j, Z$ and $ \it{l}$  prior to the limit, one gets
\be
W_{0} = B(U)\hat{Z}/{\hat{\it{l}}} + B_{j}(U)X^{j}/{\hat{\it{l}}} + w(U),
\ee
with  $ B(U)= K^{M}B_{M}$ . Notice that (12) is also  obtained by assuming that $\bar{g}_{MN}$ is not a homogeneous function of degree
zero in $Y^j, Z$ and $\it{l}$. Provided $B(U) \neq 0$, it is possible to set $B(U) = 1$ in (11) by redefining the coordinate $U$ and
the metric functions $B_{j}$, $w$, $C_{jk}$ and $ \lambda$. From now on we shall assume that $B(U) \neq 0$ in the limit and set $B(U) =
1$.

  It will be convenient to display the limiting metric also  in the harmonic
coordinates
 $\{u, v, x^j, z\}$ which covers the whole of the plane wave manifold. The transformation between these two coordinates
is well-known\cite{kn:guv}  and (11) can be cast into the form
\be
d{\hat{s}}^2 = \frac{\it{l}^2}{[z/\lambda +b_k(u)x^k + \it{l} c(u)]^2}[2 dudv - h_{ij}(u) x^i x^j du^2 - \delta_{ij}dx^i dx^j -(dz
-
\gamma du)^2],                                                                                          
\ee
where $b_{k}(u), c(u)$  and $h_{ij}(u)$ are arbitrary functions of $u$ and $\delta_{ij}$ is the Kronecker delta. We have dropped the
hats on the parameter $\it{l}$ for notational convenience. The function  $ \gamma $  depends linearly on all the
transverse coordinates $x^{j}, z$ and encodes all the
information about the conformal Killing vector $ K^M =  \lambda(u) {\delta^M}_z$ :
\be
\gamma = \lambda \dot{\omega}_{j}x^j + \dot{\lambda}z/\lambda.
\ee
Here $\omega_{j}(u)$ characterizes the twist of $K^M$; the conformal Killing vector will be 
hypersurface orthogonal if $\dot{\omega}_j = 0$. A dot over a quantity denotes differentiation with respect to $u$.

We have thus seen that any spacetime which admits a spacelike conformal Killing vector has a limit which is conformal to plane waves.
The precise from of the conformal factor that one gets depends, of course, on the homogeneity assumptions about $B_{M} (X^{L})$ and $ 
w(X^{L})$, and (13) turns out to be suitable for Einstein spaces. Let us consider on $M_{D}$  the Einstein-Hilbert action
\be
S_{EH} = - \frac{1}{2\kappa^{2}_{D}}\int{\: d^{D}X}\: \sqrt{-g} (R - 2\Lambda),
\ee 
where $\kappa_D$ is the gravitational constant, $g$ is the determinant of $g_{MN}$, $ R$ is the curvature scalar and $\Lambda$
is the cosmological constant. When   $\Lambda$ is parametrized as
\be
\Lambda = \frac{-n(n-1)}{2{\it{l}}^2},
\ee
so that  $\Lambda \leq 0$  and the metric, the cosmological constant and the coordinates  are rescaled according to (8)-(10),  the
Einstein-Hilbert 
action transforms as
\be
 \hat{S}_{EH} = ( \kappa^{2}_{D} / {\hat{\kappa}^{2}_{D}}) \Omega^{(2-2D)} S_{EH}.
\ee
Here we have taken into account a possible rescaling of the gravitational coupling constant $\kappa_{D}$ by introducing
$\hat{\kappa}_{D}$ and assumed that the limits of integrations are independent of $\Omega$. It can be deduced from (18) that if one
starts with a metric $g_{MN}$ which is a solution of the Einstein equations
\be
G_{MN} = - \Lambda g_{MN}, 
\ee
where $\Lambda$ is given by (17),
then its limit will be again a solution of the same set of equations  with $ \hat{\Lambda} \leq 0$. Provided that 
$\Lambda  \neq 0$, (19) imposes the following conditions on the
metric
functions of (14):
\be
\lambda^{-2} + b_{j}b_{j} =1,
\ee
\be
\ddot{c} = 0,
\ee
\be
2\dot{b}_j = -\dot{\omega}_j,
\ee 
\be
\ddot{b}_{j} = h_{jk}\:b_k,
\ee
and
\be
h_{jj}=  - \ddot{\lambda}/\lambda - {\lambda}^2 \dot{\omega}_j \dot{\omega}_j /2,
\ee
where the repeated indices are summed with $\delta_{ij}$.

      When $\it{l} \rightarrow \infty$, the cosmological constant vanishes and the original Penrose limits of the vacuum solutions are
regained. In this case the function $c(u)$ just represents the conformal invariance of the family of
vacuum plane waves and after setting $c=1$, field equations reduce only to (24). On the other hand, if $\it{l} \neq \infty$  and
one specializes to $D=4$, one finds that
the Weyl tensor of (14) vanishes when (20) - (24) are imposed. Consequently,
(14) is just the AdS metric  expressed in a curvilinear coordinate system if $ D = 4$. This is not
suprising because, the only $\Lambda < 0$ spacetime
which admits a proper conformal Killing vector in $D=4$ is the AdS spacetime \cite{kn:gar}. This means that  AdS
spacetime is closed under the above  limit and not much is really gained over the Penrose limit in $D=4$. The procedure becomes more
interesting if $D > 4$ in which case a large class of Einstein
spaces admits conformal Killing vectors.  Under rather mild assumptions  one can show that $D > 4$,  $\Lambda \neq 0$  Einstein
manifolds which admit conformal Killing vectors must have locally the warped product structure:
\be
M_D = M_{d} \times_{f} N_{D-d} ,
\ee
where $M_d$ is a d-dimensional space of constant sectional curvature $k$  and $N_{D-d}$ is an Einstein manifold \cite{kn:ker}.
Let $\bar{x}^a$ and $y^{\Gamma}$ be the coordinates and $g_{ab}$ and $g_{\Pi \Gamma}$ be  the metrics on $ M_d$ and $N_{D-d}$
respectively. In
this coordinate
system, which is not adapted to $ K^M$, the line element (3) can be written as
\be
ds^2 = g_{ab}(\bar{x}^c)d\bar{x}^a d\bar{x}^b + f^2(\bar{x}^c) g_{\Pi \Gamma}(y^{\Theta}) dy^{\Pi} dy^{\Gamma},
\ee
where the gradient of the warp function $f$ is a closed conformal vector field on $M_d$ satisfying
\be
k f^2 - g^{ab} \nabla_{a}f \nabla_{b}f = m,
\ee
and $k = -1/{\it{l}^2}$. The Ricci tensor of $ N_{D-d}$  satisfies $ R_{\Pi \Gamma} = m(n - d)g_{\Pi \Gamma}$, where $m$ is a constant
but
$M_d$ will be an Einstein space if and only if it is conformally flat \cite{kn:ker}.
 The metrics on the limits of these warped product manifolds are given by (14) and governed by (20) - (24).

The existence of a conformal Killing vector on $M_D$ as well as the hypersurface orthogonality of $K^M$ are both hereditary properties
in the sense of \cite{kn:ger}.
If $K^M$ is assumed to be hypersurface orthogonal, (22) requires that $\dot{b}_{j}(u) =0$ and (23) reduces to
\be
h_{jk}(u) b_k = 0.
\ee
Whenever $h_{jk}$ is an invertible matrix, it follows that  $b_j =0$ and $\lambda^2 =1$. In such cases $c(u)$ can be set equal
to zero by redefining the coordinates $z$ and  $v$ provided $\it{l} \neq \infty$. Under these assumptions  (14) becomes
\be
ds^2 = \frac{\it{l}^2}{ z^2} [ 2du dv - h_{ij}(u) x^i x^j du^2 - \delta_{ij} dx^i dx^j - dz^2 ].
\ee 
The only implication of (19) is now
\be
h_{jj}(u) = 0,
\ee
and one gets precisely the nonlinear version of the Randall-Sundrum zero mode which was discussed in \cite{kn:cg}.

 It is therefore clear that any Einstein space  $M_D$ having the metric
\be
ds^2 = \frac{\it{l}^2}{ Z^2} [ \bar{g}_{\mu \nu} (x^{\alpha}) dx^{\mu} dx^{\nu} - dZ^2],
\ee
where $\bar{g}_{\mu \nu}$ is Ricci-flat, will have a limit that is given by (29) and (30). Such metrics have been considered
 in
the brane world scenarios where a domain wall $\Sigma$, located at $ Z = \it{l}$, splits  $M_D$ into two regions $M_{D}^{\pm}$. In such 
frameworks the above discussion needs to be refined in two respects.
 First of all, since the normal derivatives of $g_{MN}$
may suffer discontinuities on $\Sigma$, one should now either assume that (1) holds in the bulk outside $\Sigma$
 or treat $\Phi$ as a distribution. Secondly, one should see whether the total action \cite{kn:cr}
\be
S = S_{EH} + S_{GH} + S_{DW},
\ee
also transforms homogenously under the rescalings (8)- (10) that are needed in the limit. The last two terms of (32)
are due to the presence of $\Sigma$ in $M_D$.  Let $n^M$ be the unit normal to $\Sigma$ and $g_{MN} = q_{MN} - n_{M}
n_{N}$ so that $q_{MN}$ is the induced metric  on $\Sigma$.
 The Gibbons-Hawking term is
\be
S_{GH} = \frac{1}{\kappa^{2}_{D}} \int_{\Sigma \pm } \:d^{n}X \sqrt{-q} K,
\ee
where $K$ is the trace of the extrinsic curvature: $K = q^{MN} K_{MN}$, $K_{MN} = q_{M}^{P} q_{N}^{Q} \nabla_{P} n_{Q}$. The last term
of (32) is the action for the
matter on the domain wall and for our purposes it will be sufficient to take
\be
S_{DW} = T \int_{\Sigma} \:d^{n}X \sqrt{-q},
\ee
where $T$ is the tension of $\Sigma$.

 When $\Sigma$ is initially  located at $Z= \it{l}$, the Einstein-Hilbert action for each of $M_{D}^{\pm}$ will be
\be
S_{EH} = -\frac{1}{2\kappa^{2}_{D}} \int \: d^{n}X \int_{\it{l}}^{\infty}\:dZ \: \sqrt{-g}\: (R - 2\Lambda),
\ee
and after choosing the appropriately directed unit normals on the two sides of $\Sigma$, the standard tuning argument \cite{kn:witz}
gives
\be
K = \frac{-n}{\it{l}}, \hspace{.7in}   T_{initial} = \frac{2(n-1)}{\kappa^{2}_{D} \it{l}},\hspace{.7in}    \kappa^{2}_{n}
=\frac{n-2}{2\it{l}} \kappa^{2}_{D},
\ee
where $\kappa_{n}$ is the n-dimensional effective gravitational constant. After the rescalings (8) and (9), $\Sigma(\Omega)$ is
positioned at $\hat{Z} = \hat{\it{l}}$ on $M_{D}^{\pm}(\Omega, \hat{\it{l}})$ and consequently, $S_{EH}$ involves the integrations
\be
S_{EH} =-\frac{\Omega^{D}}{2\kappa^{2}_{D}} \int \: d^{n}X_{r} \int_{\hat{\it{l}}}^{\infty} \: d\hat{Z}\: \sqrt{-g}\: (R - 2\Lambda),
\ee
where $X_{r}$ stands for the rescaled coordinates. The extrinsic curvature is now $K(\Omega)= -n/\Omega \hat{\it{l}}$. If one then
performs the conformal transformation (10), $S_{EH}$ picks up another $\Omega^{(D-2)}$ factor and $K$ also transforms: 
$K(\Omega) \rightarrow \Omega
K(\Omega)$. Therefore, $\hat{K} =-n/\hat{\it{l}}$. Since $S_{GH}$ and $S_{DW}$ both involve $(D-1)$-dimensional volume elements, the
endpoint
of (8) - (10) is a homogeneous transformation of the total action:
\be
S = \Omega^{(2D - 2)} \hat{S}(\kappa_{D},\hat{g}_{MN}, \hat{\it{l}}),  
\ee
which means that the cosmological constant and the tension can be tuned on each of $ M_{D}(\Omega,\hat{\it{l}})$, including the boundary
$\Omega\rightarrow 0$, and one finds 
\be
\hat{K}=\frac{-n}{\hat{\it{l}}}, \hspace{.7in} T=\frac{2(n-1)}{\kappa^{2}_{D}\hat{\it{l}}},\hspace{.7in}
\hat{\kappa}^{2}_{n}=\frac{n-2}{2\hat{\it{l}}} \kappa^{2}_{D}.
\ee
 One may therefore conclude that
the Randall-Sundrum zero mode is indeed the Penrose limit of any brane world scenario which is based on (31) and (32). Notice that in
the limiting spacetime
\be
\hat{\kappa}_{n} = \Omega \kappa_{n}, \hspace{.7in} T=\Omega T_{initial},
\ee
and one is dealing with the large tension and large $\kappa_{n}$ limit of the original domain wall while keeping $\kappa_{D}$ fixed.

Finally, let us note that the above discussion cannot be carried over directly to Einstein spaces which have $\Lambda > 0$. This is
because, although the Einstein-Hilbert Lagrangian still transforms homogeneously
under (9) and (10) when the sign of $\Lambda$ is changed in (17), the above gauge choice is no longer available. If one demands to
include
the 
 de Sitter (dS) space  in the limits of such spacetimes,
 the conformal factor $W(X^L)$ must depend linearly on a homogeneous time coordinate:  $X^{0} = (Y^{+} + Y^{-})/\it{l} $, but then
(7) will not be 
a suitable gauge choice. Therefore, the procedure breaks down when $\Lambda > 0$  and it is not possible to
obtain an analog of the AdS plane waves in a dS space by the same analysis. This is consistent with the $ D=4$ result that the
only Einstein spaces that are properly conformal to non-flat pp- waves are the Siklos solutions \cite{kn:sl}.

\section*{Appendix}
Our conventions are as follows: In all $D \geq 4$ we use the ``mostly
minus''
signature $ (+,-, \ldots, -)$ and the orientation $ \epsilon_{012 \ldots
D-1} = 1$. The Ricci tensor is defined as $R_{M N} = {R^K}_{MNK}$ and the Riemann curvature obeys $ 
(\nabla_N \nabla_ M -
\nabla_M \nabla_N) T_K = {R^L}_{KMN} T_L $
for an arbitrary vector field $ T_M$.  
   
\section*{Acknowledgements}

 The research reported in this paper has been
supported in part by the
Turkish
Academy of Sciences(T{\"U}BA).

\end{document}